\documentclass[9pt,twocolumn,twoside]{pnas-new}

\templatetype{pnasresearcharticle} 

\title{Template for preparing your research report submission to PNAS using Overleaf}

\author[a,b,c]{Gerd Graßhoff}
\author[a,b]{Gordon Fischer} 

\affil[a]{Humboldt University Berlin}
\affil[b]{Excellence Cluster Topoi}
\affil[c]{Max Planck Institute for the History of Science}

\leadauthor{Graßhoff} 

\significancestatement{
\begin{enumerate}
\item Five hundred years ago, Copernicus designed a new type of astronomical measuring instrument, based on the principles of reflecting and inclining vertical sundials.
\item The instrument's lattice of lines above the doorway to Copernicus's rooms at Olsztyn Castle were constructed geometrically and not plotted empirically. 
\item The reflected light of the Sun passes the three major vertical lines on the right side of the panel every equatorial hour.
\item  We were unable to find an hour line for local noon in the historical records or during the instrument's recent restoration.
\item The spot of sunlight passes the noon hour line half an equatorial hour past local noon.
\end{enumerate}
}

\authorcontributions{
Gerd Graßhoff was responsible for the overall research project, the history of science and data acquisition. Gordon Fischer wrote the Python program and was in charge of data organisation.
This project was supported by the Cluster of
Excellence 264 TOPOI, research group D5.  The photographs of
Copernicus's heliograph were taken by Gerd Graßhoff and Joanna
Pruszyńska, with the kind permission of the Museum of Warmia and
Masuria in Olsztyn, Poland (director E. Jelińska). Bernhard Fritsch
processed the photographs for 3D models, using structure-from-motion
software Photoscan. We are grateful for the valuable discussions held with
Noel M. Swerdlow and Richard L. Kremer.
}

\correspondingauthor{\textsuperscript{2} E-mail: gerd.grasshoff@hu-berlin.de}
\begin{abstract}
Exactly 500 years ago, Nicolaus Copernicus drew a lattice of lines on
a panel above the doorway to his rooms at Olsztyn Castle, then in the
Bishopric of Warmia. Although its design has long been regarded as some kind of reflecting vertical sundial, the exact astronomical designation of the lines and related measuring techniques remained unknown. Surprisingly, Copernicus did not refer to his new observational methods in his principal work, \textit{De Revolutionibus}. A data analysis of a 3D
model of the panel has, at last, solved the mystery: Copernicus
created a new type of measuring device -- a heliograph with a non-local
reference meridian -- to precisely measure ecliptic longitudes of the
Sun around the time of the equinoxes. The data, 3D model and modeling
results of our analysis are open access and available in the form of
digital (Jupyter) notebooks.
\end{abstract}

\dates{This manuscript was compiled on \today.}
\title{Copernicus's Heliograph at Olsztyn -- the 500th Anniversary of
a Scientific Milestone}

\begin{document}
\maketitle
\thispagestyle{firststyle}
\ifthenelse{\boolean{shortarticle}}{\ifthenelse{\boolean{singlecolumn}}{\abscontentformatted}{\abscontent}}{}

\section{From Frombork to Olsztyn}

In November 1516 Nicolaus Copernicus, canon of the chapter of Frombork
(in modern-day Poland), was unexpectedly elected to the office of
chief administrator of the southern district of Warmia, based at
Olsztyn Castle, as the previous incumbent had fallen ill.
%

Since 1512 Copernicus had been working on his revolutionary
heliocentric theory of planetary motion in Frombork, where he had
established a small observatory. Of the instruments used there, only
three are mentioned in his principal work, \textit{De Revolutionibus}:
a parallactic triangle, which was accurate and easy to handle; a
quadrant for measuring the height of celestial objects at culmination;
and an armillary sphere, used, albeit rarely, for other more general
astronomical configurations (cf. \cite{dobrzycki_1975}, p. 3).
%
 
\begin{figure*} \centering
\includegraphics[width=1\linewidth]{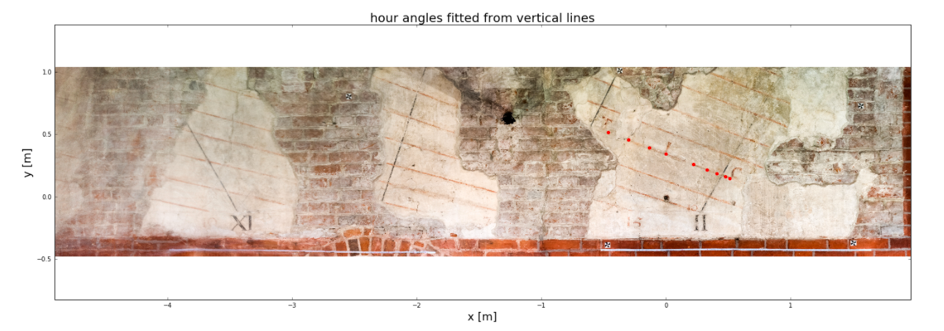}
	\caption{Orthophoto taken from the 3D model of
Copernicus's heliograph, which is located above the outer doorway to his rooms at
Olsztyn Castle.}
	\label{fig:screenshot001}
\end{figure*}

While in Frombork but before being sent to Olsztyn, Copernicus became involved in the reform of the calendar through the commission of the Fifth Council of the Lateran (1512--17) over which Paul of Middelburg presided. Both the pope and the emperor had asked the leading astronomers of Europe, including Copernicus, to provide astronomical justification for reforming the Julian calendar that would be based on a revised length of the tropical (or solar) year. The heliograph of Copernicus was not, however, the astronomer's answer to this request; the commission failed to come to any agreement in the
concluding discussions held in 1516 \cite{minnich1969} and by the time Copernicus had designed the heliograph at Olsztyn, the subject had been dropped. Copernicus, meanwhile, focused his attention on understanding the causes of the irregularities of the motion of the Earth, expressed in terms of the variations in the
length of the tropical year.

On moving to Olsztyn, Copernicus most probably had to leave his
astronomical instruments behind. Nevertheless, he continued his
astronomical studies and created a new measuring tool: Copernicus
carefully plastered an area above the outer doorway to his rooms in
the castle's gallery and drew a lattice of lines on the wall, the
remaining part of which measures 7.05m x 1.4m (cf.
\cite{szubiakowskijacek_2013}, p. 62,
\href{https://doi.org/10.17171/2-7-17-1}{3D Model:
  doi.org/10.17171/2-7-17-1}). This panel has survived largely intact
(Fig. \ref{fig:screenshot001}). For an explanation of how to access the data notebooks, please refer to this \href{https://doi.org/10.17171/2-7-20-1}{instruction video
doi.org/10.17171/2-7-20-1.}

Because of the design of the lines and the now fragmentary
inscriptions that accompany them, we can confirm the received view that this panel uses the principles of a reflecting and inclining vertical sundial: the
red horizontal lines, which slope slightly downwards to the right, are
called day lines. On a sundial these lines would represent the daily
movement of the reflected spot of light on the wall from west to east
(from left to right on our orthophoto). One of the lines bears an
inscription, of which only a few letters have survived; it was
tentatively emended by Przypkowski to ``AEQUINOCTIUM''. Other readings are, however, possible. For example, it was long considered to represent the 
equinoctial line for the motion of the Sun during the equinoxes. 
%
The black vertical lines, which slope upwards to the right, show the hours of the day according to the position of the spot of
light on the panel.

\begin{figure} \centering
\includegraphics[width=0.8\linewidth]{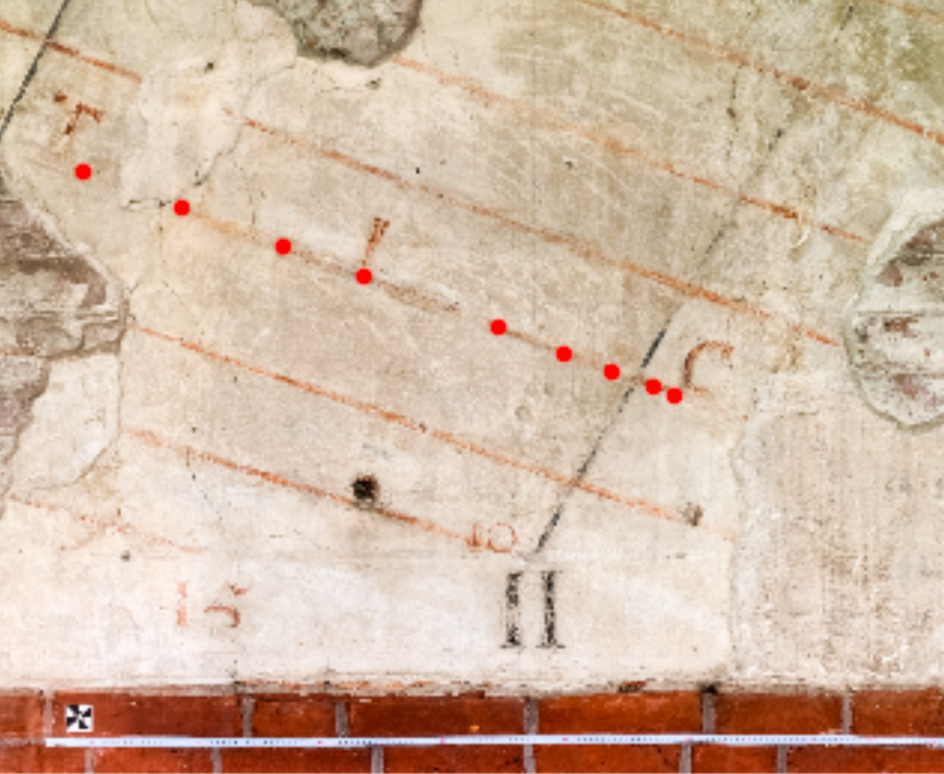}
	\caption{Detail of Copernicus's heliograph at hour line
``II". The red dots mark the path of the reflected light of the Sun
during the day before 2 p.m. at the equinoxes.}
%
	\label{fig:screenshot002}
\end{figure}

The orthophoto (Fig. \ref{fig:screenshot001}) shows horizontal,
slightly downward-inclining day lines for the ecliptic longitudes of
the Sun every five degrees. During the day, the projected spot of
light of the Sun runs parallel to these lines, from left to
right. Four marked vertically oriented lines are visible, which we
interpret as hour lines: on the far left, the eleventh hour line
labelled ``XI"; next what we call the ``noon line'' -- its label is lost; then, the fragment of the unmarked first hour line; and finally the hour line marked ``II" on
the far right. A detailed orthophoto
(Fig. \ref{fig:screenshot002}) shows the lines at the second hour
line, with the red dots representing the simulated motion of the spots
of light at the equinoxes. The small red numbers ``10" and ``15" possibly
%
indicate the path of the spots of light of the respective degrees of solar
longitudes. The aligned letters above the red dots mark the
equinoctial line.

An astronomical instrument of this size and construction had not
existed before Copernicus. None of the astronomer's documents
or published texts mentions the panel, its \textit{modus operandi}, or
the data that it measured.  Arguably, no other astronomer before Copernicus ever used such an instrument for measuring solar longitudes. In 1654, the French Jesuit Pierre Gassendi wrote an unflattering description of the panel: 
``In the castle of Allenstein [the German name of Olsztyn], a dozen lines on a wall can be seen which are supposed to have been drawn by
Copernicus, who would have represented a kind of sundial, but this
mural drawing looks way too primitive and awkward to be of any use in
astronomy'' (cf. \cite{gassendi_2002}, p. 124). The panel soon came to
be regarded as a curious oddity of little or no astronomical
significance, and historians of science soon lost interest in
it. Nonetheless, the panel was recognised as a Copernican artefact; it
was reproduced in Zinner \cite{zinnere_1956} without any deeper analysis.
The panel has undergone several restorations, particularly after the
plaster was partially removed during its restoration in
1954/56. Przypkowski believed that the horizontal lines represent the day lines
of a reflecting sundial
(\cite{przypkowskit_1973},\cite{przypkowskitadeusz_1960}), although he was not able to find an explanation for the vertical lines. Ten years ago the panel
was carefully restored again and critically examined at the Museum of
Warmia and Masuria (Muzeum Warmii i Mazur) at Olsztyn Castle. By analysing the paint and the plaster, the restorers were able to date the visible lines 
to Copernicus's time. In 2013,
the museum published an excellent summary of the astronomical panel's
restoration work, which includes a photogrammetric assessment
\cite{jelinska_2013} and interpretation of the panel.  GIS data and
photogrammetric measurements were supplied by Mialdun, which we have used in this study
(\cite{mialdunjerzy_2013}, p. 36), and Szubiakowski provided a helpful
discussion of the panel's astronomy
\cite{szubiakowskijacek_2013} as well as his own analysis of the astronomical table. Both Miałdun and Szubiakowski analysed the
geometry of the lines: they were able to fit the day lines and thereby
determine the approximate position of the reflecting mirror. However,
the meaning of the hour lines remained unresolved; there was still no agreement on how to
interpret Copernicus's table.  Therefore, on the Olsztyn panel's 500th anniversary, it seems particularly appropriate to
re-examine its purpose.

\begin{figure} \centering
\includegraphics[width=0.5\textwidth]{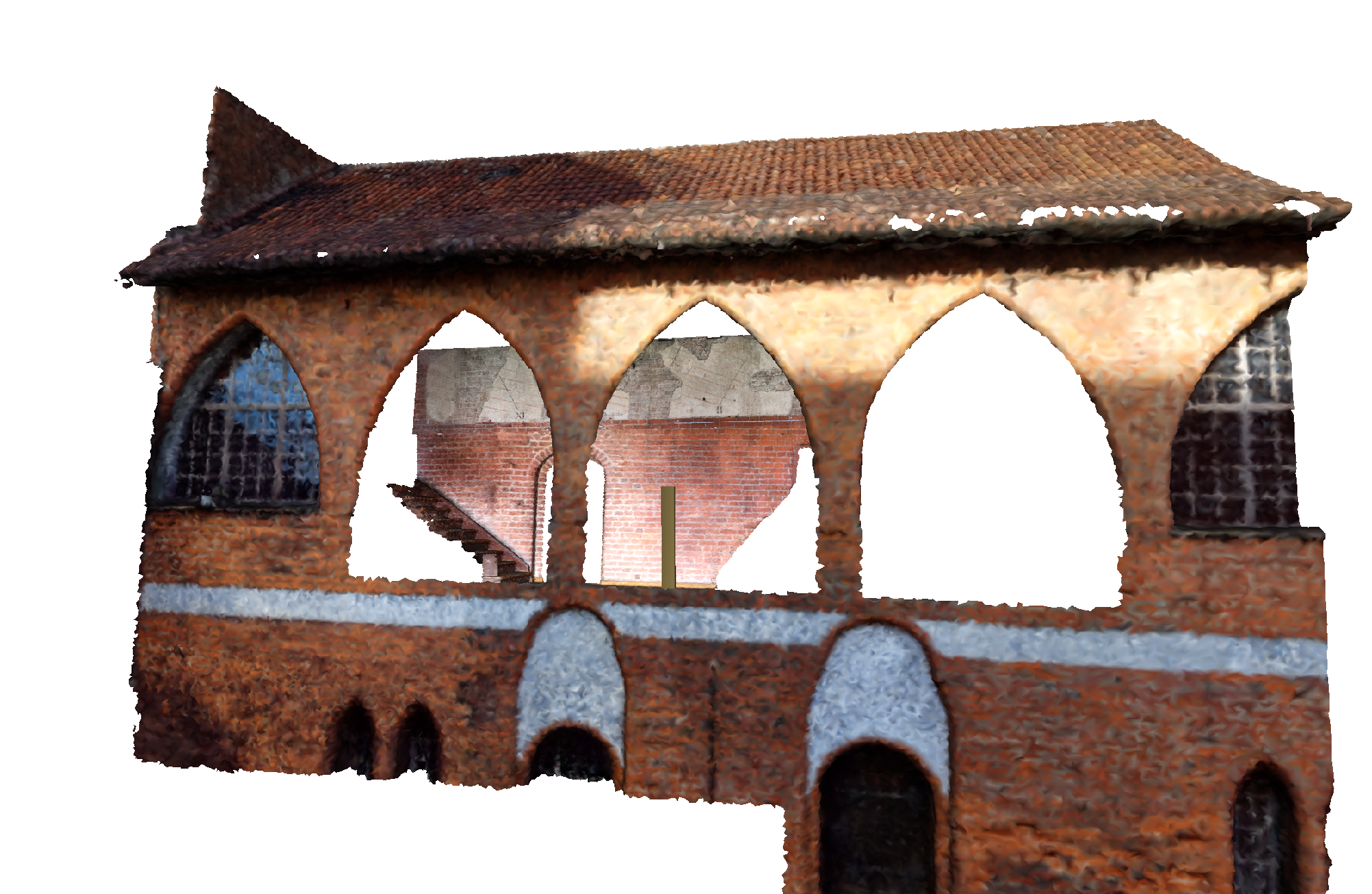}
	\caption{3D model of Copernicus's heliograph at Olsztyn
Castle. The reflecting mirror has been placed in the model and is visible through the central window in front of the wall
\href{http://repository.edition-topoi.org/collection/COPS/single/00012/0}{(doi.org/10.17171/2-7-17-1).}}
	\label{fig:intro}
\end{figure}

\section{Copernicus's measuring lattice}

To set up the model we followed the received view that the
panel lines were constructed according to the principles of a reflecting vertical sundial with a horizontally oriented mirror that reflects the incident sunlight onto a vertical
wall. During the day the projected spot of light moves in the opposite direction of the
%
diurnal motion of the Sun across the wall from west to east. At
the equinoxes, on a wall aligned exactly in an east--west direction, the
%
spot of light would move during the day along a straight horizontal line from west to east. The height of this horizontal day line depends solely on the
mirror's distance from the wall and the geographical latitude of the
instrument. If the wall turns away from its north direction (as at Olsztyn
%
%
Castle), the equinoctial lines will be straight but tilted.
%

\begin{figure} \centering
\includegraphics[width=0.9\linewidth]{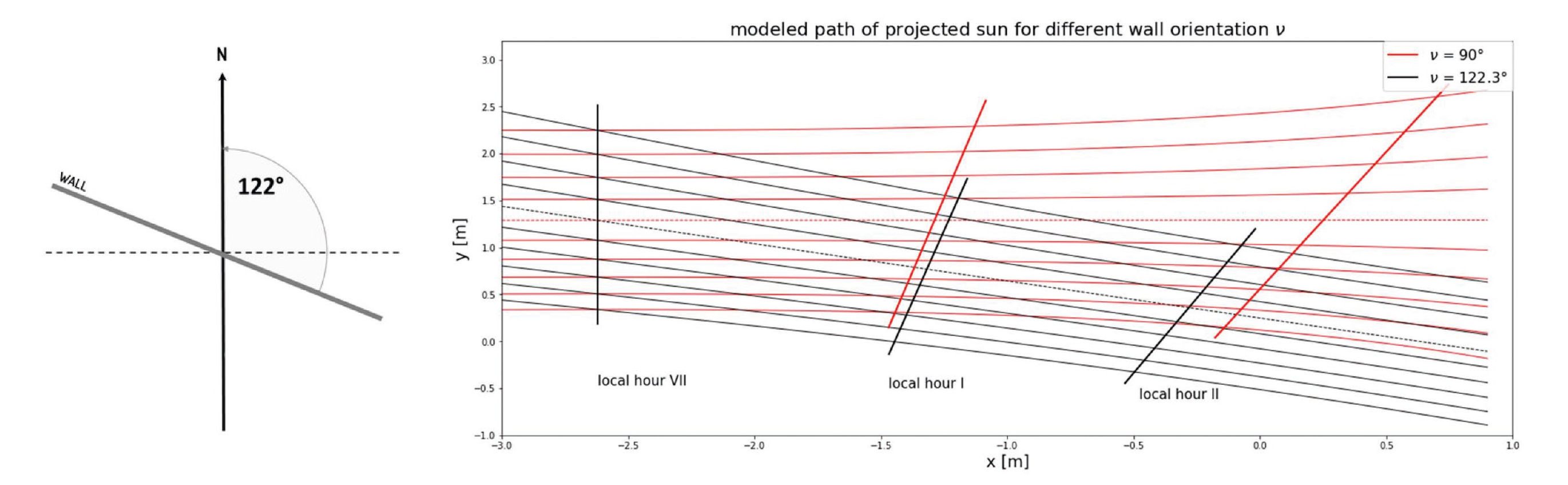}
	\caption{Plots of day lines for the motion of the projected
spot of light at five-day intervals for different orientations of a
wall. The black lines track the daily motion on a wall oriented exactly
west to east; the red lines mark the daily motion on a wall inclined
to the north, as at Olsztyn Castle. The noon hour
line (local hour VII) on the wall is vertical for all rotations. (See
\href{https://doi.org/10.17171/2-7-18-1}{doi.org/10.17171/2-7-18-1.)}}
	\label{neig}
\end{figure}

If one plots the apparent daily motion of the Sun on the wall on
different days, one obtains a group of hyperbolas for the day
lines on the wall. For the part of the wall that Copernicus used, the
hyperbolas are closely approximated by straight lines. 
Fig. \ref{neig} simulates the paths of the projected spot
of light at five-day intervals: the black lines show the paths on a
wall with an exact east--west alignment; the red lines show the paths
on a wall tilted about 30°, the angle of the
wall at Olsztyn Castle measured using the GIS data by J. Miałdun
(cf. \cite{mialdunjerzy_2013}, p. 35ff.).

The straight day line in the center tracks the spot of light during
%
the spring and autumn equinoxes. Its inclination closely fits the
model for a wall of that geographical orientation, and strongly supports the view that Copernicus
constructed the panel using the principles of a reflecting
sundial. However, it is highly unlikely that Copernicus designed and
used the panel as a sundial: it is simply too large, it would have
only shown a fraction of the day, and it does not have the hour lines for the usual time system. 
%

The horizontal lines seem to show the path of the spot of light at intervals of five days or five degrees of specific 
ecliptic longitudes of the Sun (or on specific days of the year). 
We pick up the thread of Dobrzycki: ``Thus the preserved remnants of a solar-observing table in Olsztyn Castle show Copernicus as the
designer of a theoretically interesting instrument. A net of
hyperbolic lines marks the path of the solar image, reflected on the
wall by a horizontal mirror; the lines are drawn for every fifth day,
for a period of approximately a month before and after the equinoxes''
(cf. \cite{dobrzycki_1975}, p. 29). There is, however, a difficult riddle:
the hour lines are simply not compatible with the standard design of a
vertical sundial showing local hours. 

\begin{figure} \centering
	\includegraphics[width=0.5\linewidth]{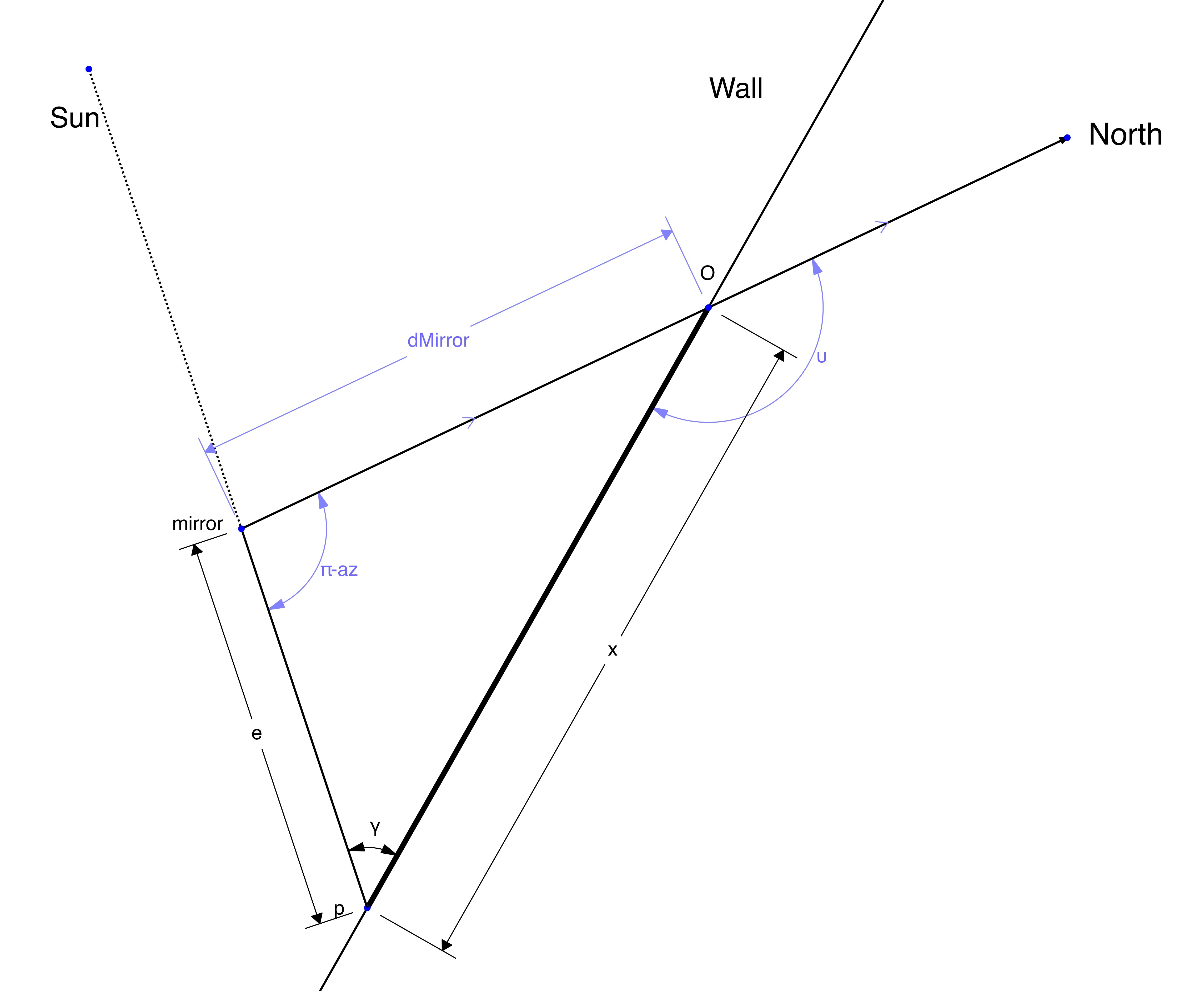}
 	\caption{A horizontal cross-section of the wall's geometry, with the reflecting mirror (left), sunlight from above and projected sun spots on Copernicus's heliograph at the wall as the right part of the triangle.}
%
\label{fig:geometry}
\end{figure}
\section{Fitting the model to data}

In September 2016 Gerd Gra{\ss}hoff and Joanna Pruszyńska took a series of
photographs of the panel. Using  structure-from-motion techniques, we
created a 3D model from the photos. The 3D model was calibrated using Miałdun's GIS data and serves as a reference template for calibrating the photos. All the detail photos were registered with the 3D model and allow for a precise computational mapping of their pixels to the panel's metric reference frame. This provides a high resolution measurement of the geometry of the heliograph that surpasses previous results.
%
%

On this basis, we studied an astronomical model of the panel's geometry
by simulating the lines, which enabled us to determine the 
position of the mirror that Copernicus had placed in front
of the wall. We also tested possible interpretations of their astronomical
meaning, especially those of the hour lines.

Reconstructing the hour lines proved particularly challenging. Fig. \ref{neig} shows the hour lines intersecting the day lines of ecliptic longitudes at five-degree intervals for the period of one month before and after the equinoxes. If these lines are plotted for local hours at hourly
intervals (as in sundials), we obtain lines that differ fundamentally
to Copernicus's lines: the noon line needs to be oriented vertically,
independent of the orientation of the wall. Copernicus's panel
clearly has no such lines -- neither on the restored panel layer from Copernicus's time, nor on later layers. This omission cannot be explained away as
an inaccuracy on the part of Copernicus when constructing the panel –
an astronomer would never have made such an outrageous error. 
Thus, it rules out that the system of lines are a reflecting
sundial with local hour lines. So was Gassendi right to have described the
panel as a rather primitively constructed sundial of no astronomical
value?

For our re-evaluation and reconstruction of the astronomical function
of the panel, we developed a multi-parametric model to calculate the
lines of the projected spots of light. The panel was modeled using
a 3D reconstruction of the wall employing current data analysis
tools. The Jupyter notebooks of data analysis can be accessed
interactively through the server environment. 

The model for the movement of the spots of light on the wall assumes a
geometrical idealisation (see Fig. \ref{fig:heliograph}). The
straight line passing through points $O$ and $P$ represents the wall
on which the lattice of lines was drawn. The wall is orientated to the
north by angle $\nu$. A horizontally calibrated mirror was placed in
front of the wall at the position marked $mirror$. From the mirror, a
meridian line runs north and intersects the wall at point
$O$. Sunlight is reflected with an azimuth direction $az$ (the
horizontal elevation from the north) and, with the height of the Sun
($alt$) above the mirror and hits the wall at point $P$. The latter is
%
located at the point on the wall where horizontal distance $x$ and
height $y$ meet $O$ above the mirror.

The model's parameters were
approximated to a best fit of the marked lines on the wall. When more lines are incorporated into the
optimisation, the panel's approximate parameters increase in significance. For this reason, it was important to include the hour lines in
the reconstruction.

\begin{figure}
	\centering
	\includegraphics[width=0.7\linewidth]{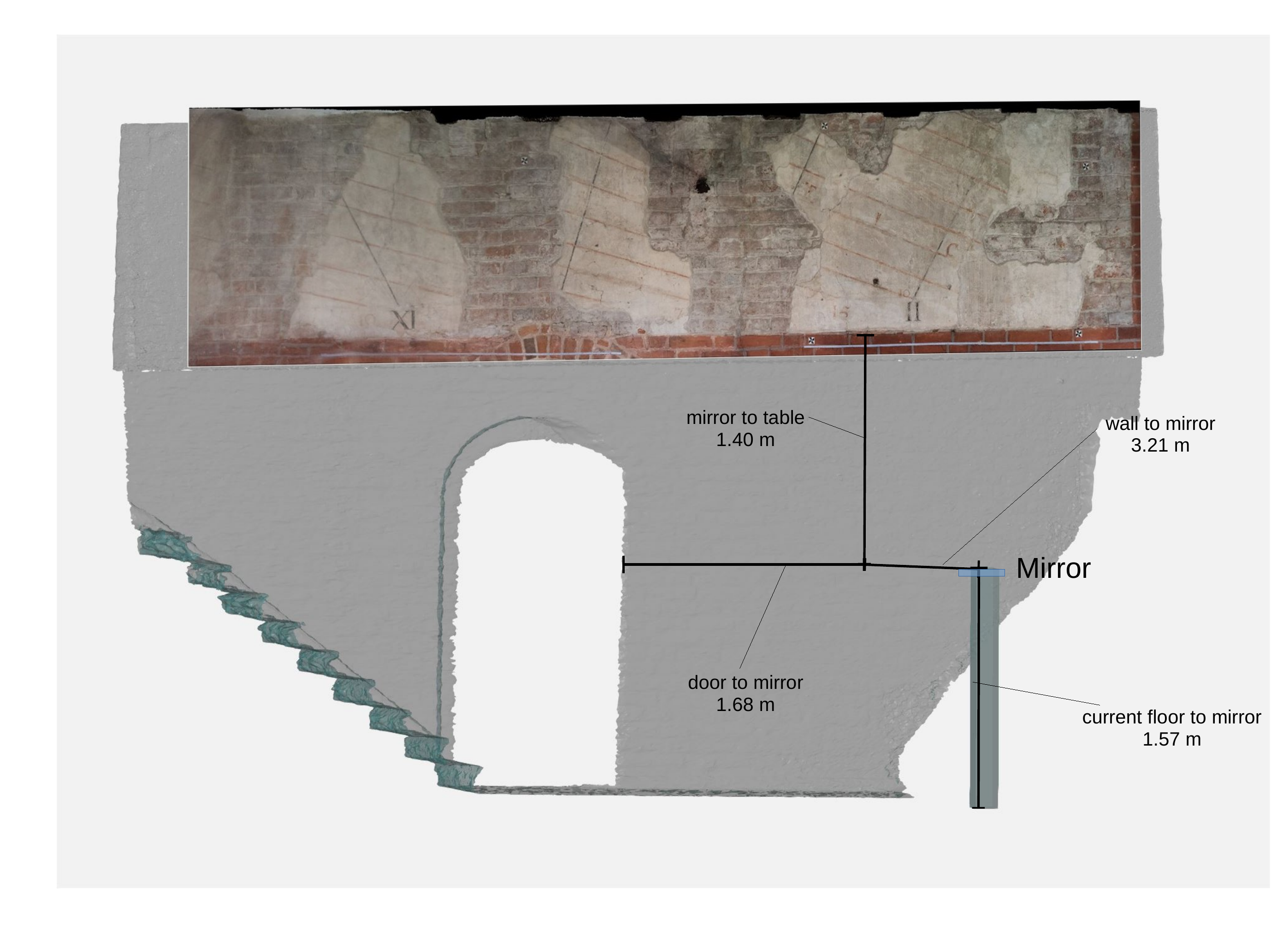}
	\caption{Geometry of the modeled position of the mirror in front of the wall using orthogonal coordinates.}
	\label{fig:mirrorPosition}
\end{figure}

\section{Modeling results}
%

The horizontal day lines on a perfect vertical sundial would be in the form of hyperbolic curves. A close analysis of the geometry of the panel's lines shows that they are not hyperbolic curves but straight lines. The geometry is a good approximation for the intended time windows around the equinoxes and is compatible with the standard construction techniques of sundials at the time of Copernicus. This finding makes it highly unlikely that the day lines were empirically plotted following the sun's path every fifth day. The weather conditions at Olsztyn would not have been  favourable for such a procedure and we cannot imagine how Copernicus would have been able to draw a linear interpolation between empirical markings under such circumstances. This means that \textit{the grid of lines must have been geometrically constructed.}

The construction procedures for vertical sundials differentiate between the hour lines at either side of the noon line. We thus took the morning hour line labeled ``XI" on the left side of the panel as our dividing line. We iterated the reconstruction for the better-preserved right side of the panel and left the confirmation of the results to the left side of the panel. The lines in this section of the panel are less well preserved as parts of the wall are missing or are severely cracked. The model's parameters are: the wall’s spatial orientation, the position of the reflecting mirror, and the dimension of the day and hour lines, including its astronomical parameters. The reconstruction model fits both day and hour lines for a robust parameter set of the geometrical configuration of Copernicus’s heliograph. \textit{The position of the mirror can be precisely located at the position derived in front of the wall (Fig. \ref{fig:mirrorPosition}).}
%
%
%
%

\begin{figure} \centering
	\includegraphics[width=1\linewidth]{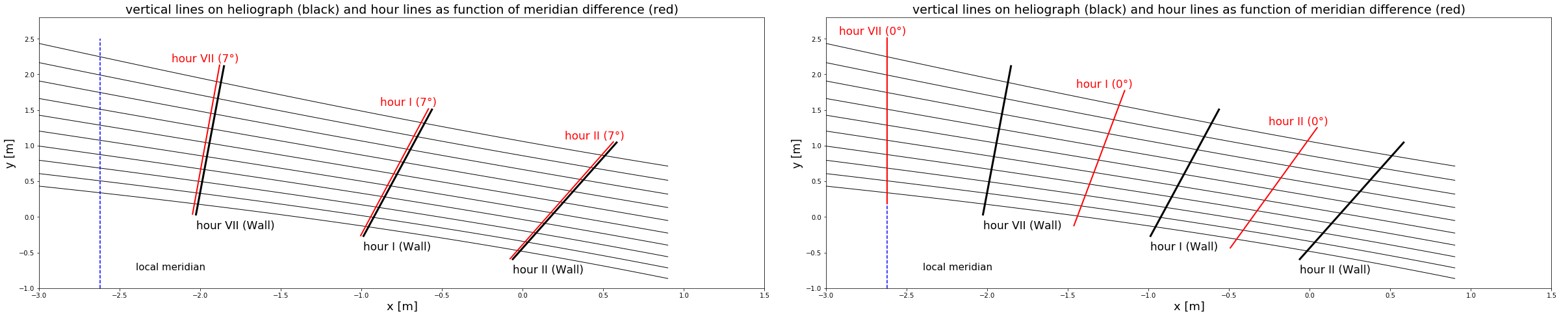}
	\caption{Difference between the hour lines for local hours (right) and the hour lines relative to a meridian shift of half an equatorial hour.}
	\label{fig:heliograph}
\end{figure}

When calculating local hours on a reflected sundial, the noon line is always 
vertical. On the heliograph, however, all the lines
are clearly tilted and shifted to the east. Either these lines were not intended to display the hours of any time system at all or Copernicus plotted the hour lines for a different reference time. 
We make no assumptions about the Roman numberings of the hour lines on the lower part of the panel in our reconstruction, which searches for a sequence of hour angles that matches the lines. In particular, we make no assumptions about a specific reference meridian. If the lines on the panel are indeed hour lines, their time difference would amount to exactly one hour between consecutive lines for all the days of the year. Each passage of the moving spot of reflected light would pass a line exactly every hour. This holds independently of the specific reference meridian and would be a characteristic criterion of an hour system used by Copernicus. It can be easily tested; indeed, all the lines on the panel are separated by 15° equatorial time and fit exactly one specific meridian difference!
%
%
An accompanying notebook analyses the effect of different reference meridians on the orientation of the hour lines for the fitted set of
%
parameters. Diagram
%
(\ref{fig:heliograph}) plots the measured lines (both day lines and
hour lines), with the above orthophoto represented as black lines. The
%
%
modeled hour lines (red lines) are computed for various meridian
differences to the local meridian and overlay the panel lines. The
interactive notebook shows the effect of meridian shifts (see the buttons at the
top) on the hour angles of Copernicus’s panel (red lines). The initial
value for Olsztyn (a difference of zero degrees) illustrates the huge deviation
%
from the drawn lines on the panel (as in Fig.
\ref{fig:heliograph}). \textit{The vertical lines are equatorial hour lines with a time difference of one hour.}
%

Remarkably, for this to be the solution the hour lines at Olsztyn would have to pass the projected spots of light every hour as needed! \textit{Fitting all the lines only requires the adjustment of one parameter: all the hour lines are shifted by half an equatorial hour.}
%
%

This solution shows a remarkably unusual concept of hours for the time of Copernicus. After the introduction of mechanical town clocks in the thirteenth century, seasonal hours began to disappear from public life \cite{Dohrn-vanRossum1996}. At the time of Copernicus it was quite common for several systems of hours, besides that of seasonal hours, to be used , as in Antiquity. Sundials were built for the system of Italian hours (counting hours from sunset), Babylonian hours (counting from sunrise) or hybrid hour systems, such as Nuremberg hours. For the Italian or Babylonian hours' system, the twelfth hour occurred when the sun was close to the horizon. Public sundials did not show equatorial hours that counted from noon as these would have been unnecessarily cumbersome for astronomical purposes. These hour systems were clearly not used in Copernicus's table, since their hour lines would shift with the seasonal rising and setting times of the Sun, which differs from the measured features of the lines on the panel. 

The solution: the panel's hour lines show equinoctial hours, 
using a reference meridian west of Olsztyn. The so-called
noon line then shows the Sun's position for a different meridian than local noon. This would mean that all the measurements made use of a standard reference time that differed from that of the local time (similar to today's Greenwich Mean Time). 

In all of the modeling scenarios the mirror was placed
4 metres from the reconstructed position of the wall's meridian
point. Two digital notebooks of the model, which proves the uniqueness of the solution, are accessible on an interactive
\href{https://doi.org/10.17171/2-7-21-1}{live server at
	doi.org/10.17171/2-7-21-1} \href{https://doi.org/10.17171/2-7-22-1}{and
	doi.org/10.17171/2-7-22-1}.

\section{Possible Copernican measurement procedures}

We have established that the panel shows geometrically constructed day lines and equatorial hour lines, and that the construction of the lines follows conventional sundial design principles. Copernicus could have taken measurements from the panel in two ways. He might have waited until the moving light spots crossed one
of the lines -- either the horizontal day line or the vertical hour
line -- although, since the panel reproduces only a small part of the
daily motion of the Sun, waiting for the spots of light to cross the
horizontal day lines would not have  provided enough data.  However, by
%
observing the passage of the spots of light through the hour lines, Copernicus
could easily have marked the point of intersection and manually measured
the distance to the reference marks for the full degrees of longitude, which
yield the Sun's corresponding ecliptic longitude
\href{http://repository.edition-topoi.org/collection/COPS/single/00019/0}{(doi.org/10.17171/2-7-19-1).}

This procedure would have enabled Copernicus to take extremely precise measurements.  Using the passage of hour lines allows time errors to be narrowed down to few minutes. The distances between the intersection of the day lines with the hour lines are of the order of several centimetres, which allow measurements of the Sun's corresponding ecliptic longitudes to be taken in fractions of degrees. The procedure's accuracy would have depended on the precision of the longitude marks on the wall. In \textit{De Revolutionibus} Copernicus published earlier equinox observations of limited accuracy that he had taken with his instruments at Frombork between 1515 and 1516. It is entirely plausible that he designed his new heliograph at Olsztyn in order to improve the accuracy of this series of measurements. However, Copernicus never published any observations, perhaps because of turbulent political circumstances, observational difficulties, or simply because the results from the panel had not fulfilled his expectations.
%

The heliograph at Olsztyn proves that not only did Copernicus revise and develop geometrical models in astronomy, but he also continued Regiomontanus's call for astronomy to be more empirically based. It also shows that Copernicus developed and used new observational techniques early on in his research, although he did not mention them in \textit{De Revolutionibus}, where he referred only to methods and instruments of Antiquity. Finally, the new reconstruction of Copernicus's heliograph at Olsztyn reveals that the astronomer of Warmia was also an empirical innovator in his field.
%
%


\end{document}